\documentclass[a4paper,aps,prd,10pt,preprintnumbers,twocolumn,superscriptaddress,nofootinbib,amsmath,amssymb]{revtex4-1}
\usepackage{graphicx}
\usepackage[utf8]{inputenc}
\usepackage[T1]{fontenc}
\usepackage{cmap}
\def\imo{i}
\def\re#1{Re(#1)}

\def\K{{\cal K}}

\begin{document}
\title{Quasinormal modes of the test fields in the novel 4D Einstein-Gauss-Bonnet-de Sitter gravity}
\author{M. S. Churilova}\email{wwrttye@gmail.com}
\affiliation{Research Centre for Theoretical Physics and Astrophysics, Institute of Physics, Silesian University in Opava, CZ-746 01 Opava, Czech Republic}
\begin{abstract}
The regularization proposed in [D.~Glavan and C.~Lin, Phys.\ Rev.\ Lett.\  {\bf 124}, 081301 (2020)] led to the black hole solutions which turned out to be the solutions of the consistent well-defined $4$-dimensional Einstein-Gauss-Bonnet theory of gravity suggested in [K.~Aoki, M.~Gorji and S.~Mukohyama, arXiv:2005.03859]. Recently the quasinormal modes of bosonic and fermionic fields for this theory were studied. Here we calculate quasinormal frequencies of the test scalar, electromagnetic and Dirac fields for the spherically symmetric black hole in the novel $4D$ Einstein-Gauss-Bonnet-de Sitter theory. The values of the quasinormal modes, calculated by the sixth order WKB method with Pad\'{e} approximants and the time-domain integration, show that both real oscillation frequency and the damping rate are suppressed by increasing of the cosmological constant. While the stability of the scalar and electromagnetic fields follows directly from the positive definiteness of the effective potential, there is no such positive definiteness for the Dirac field. Here, with the help of the time domain integration, taking into account all the modes, we prove stability of the Dirac field in $4D$ Einstein-Gauss-Bonnet-de Sitter theory.
\end{abstract}
\maketitle

\section{Introduction}

Modern experiments \cite{alternative1}, observing the quasinormal modes of the black holes or other compact objects, do not exclude alternative theories of gravity \cite{alternative2}, which attempt to solve fundamental problems such as hierarchy problem, singularity problem or quantum gravitational theory. The theories including higher order curvature corrections to the Einstein term, such as the Einstein-Gauss-Bonnet theory (for the second order) and Lovelock theory (for the higher orders), are among the most promising ones in this regard. According to the Lovelock theorem the quadratic in curvature Gauss-Bonnet term produces corrections to the Einstein equations only in higher than four dimensional spacetimes. Nevertheless it proved possible to introduce the non-trivial four-dimensional Gauss-Bonnet corrections together with the Pontryagin term in asymptotically anti-de Sitter spacetime \cite{Araneda:2016iiy,Miskovic:2009bm}.

Recently a non-trivial four-dimensional Einstein-Gauss-Bonnet theory of gravity has been proposed \cite{Glavan:2019inb}, which allowed one to avoid vanishing of the Gauss-Bonnet term contribution to the dynamics. This theory evoked a great interest and was investigated in many aspects in a strikingly short period of time \cite{Konoplya:2020bxa,Guo:2020zmf,Fernandes:2020rpa,Casalino:2020kbt,Wei:2020ght,Konoplya:2020qqh,Hegde:2020xlv,Kumar:2020owy,Ghosh:2020vpc,Doneva:2020ped,
Zhang:2020qew,Konoplya:2020ibi,Singh:2020xju,Ghosh:2020syx,Konoplya:2020juj,Kumar:2020uyz,Zhang:2020qam,Bahamonde:2020vfj,
HosseiniMansoori:2020yfj,Roy:2020dyy,Wei:2020poh,Singh:2020nwo,Konoplya:2020der,
Churilova:2020aca,Kumar:2020xvu,Mishra:2020gce,Li:2020tlo,Heydari-Fard:2020sib,Jin:2020emq,Konoplya:2020cbv,
Zhang:2020sjh,EslamPanah:2020hoj,NaveenaKumara:2020rmi,Aragon:2020qdc,Malafarina:2020pvl,Yang:2020czk,Cuyubamba:2020moe}. At the same time, in a number of works \cite{Lu:2020iav,Kobayashi:2020wqy,Ai:2020peo,Fernandes:2020nbq,Mahapatra:2020rds,Tian:2020nzb,Shu:2020cjw,
Arrechea:2020evj,Ma:2020ufk,Hennigar:2020lsl,Gurses:2020ofy,Bonifacio:2020vbk} it was shown that the regularization scheme used in \cite{Glavan:2019inb} did not lead to a covariant novel four-dimensional Einstein-Gauss-Bonnet gravity with two degrees of freedom, in agreement with Lovelock theorem. Nevertheless, in \cite{Aoki:2020lig} there was formulated a consistent 4D Einstein-Gauss-Bonnet theory with two degrees of freedom, violating the temporal diffeomorphism invariance, and the cosmological solution of this theory was studied in \cite{Aoki:2020iwm}. Moreover, it was stated in \cite{Aoki:2020lig} that the black hole solutions found in \cite{Glavan:2019inb} are solutions of this well-defined theory.

Among the papers dealing with this consistent 4D Einstein-Gauss-Bonnet theory are those studying the quasinormal modes of bosonic \cite{Konoplya:2020bxa} and fermionic \cite{Churilova:2020aca} fields for the spherically symmetric asymptotically flat black holes. We extend these results to the novel four-dimensional Einstein-Gauss-Bonnet-de Sitter theory of gravity via studying of the quasinormal modes for the test scalar, electromagnetic and Dirac fields. The calculations of the quasinormal frequencies are accomplished in the stability region of the Gauss-Bonnet coupling constant pointed out in \cite{Konoplya:2020bxa}.

While the stability of the scalar and electromagnetic fields follows from the positive definiteness of the effective potential, this is not so for the Dirac field. In the Schwarzschild case one of two iso-spectral potentials is still positive definite and therefore guarantees stability of the Dirac field for other chirality. In the Schwarzschild-de Sitter spacetime both effective potentials have negative gaps and the proof of the Dirac field stability is non-trivial \cite{LopezOrtega:2012hx,Konoplya:2020zso}. Here, with the help of the time domain integration, taking into account contributions of all the quasinormal modes, we prove stability of the Dirac field in the novel four-dimensional Einstein-Gauss-Bonnet-de Sitter theory of gravity.

The paper is organized as follows. In sec. II we present the black hole metric in the novel four-dimensional Einstein-Gauss-Bonnet-de Sitter theory. In sec. III we discuss the master wave equations. Section IV is devoted to quasinormal modes of the test scalar, electromagnetic and Dirac fields. In Conclusions we summarize the obtained results and mention some open questions.

\section{The black hole metric}

The Einstein-Gauss-Bonnet-de Sitter gravity in the $D$-dimensional
spacetime can be described by the action
\begin{equation}
S=\frac{1}{16\pi}\int d^{D}x\sqrt{-g}\left[R+2\Lambda+\frac{\alpha}{D-4}\mathcal{G}^{2}\right],\label{eq:action}
\end{equation}
where $g$ is the determinant of the metric $g_{\mu\nu}$, $R$ is the Ricci scalar
and $\Lambda$ is the positive cosmological constant. The Gauss-Bonnet invariant is
\begin{equation}
\mathcal{G}^{2}=R^{2}-4R_{\mu\nu}R^{\mu\nu}+R_{\mu\nu\alpha\beta}R^{\mu\nu\alpha\beta}=\frac{1}{4}\delta_{\rho\sigma\gamma\delta}^{\mu\nu\alpha\beta}R_{\ \ \mu\nu}^{\rho\sigma}R_{\ \ \alpha\beta}^{\gamma\delta}
\end{equation}
and the Gauss-Bonnet coupling constant
$\alpha$ is rescaled  by the factor $\frac{1}{D-4}$.

The corresponding equation
of motion has the form
\begin{equation}
G_{\mu\nu}+\frac{\alpha}{D-4}H_{\mu\nu}=\Lambda g_{\mu\nu},\label{eq:EinsteinEq}
\end{equation}
where $G_{\mu\nu}$ is the Einstein tensor and $H_{\mu\nu}$ is the contribution of
the Gauss-Bonnet term.

It was suggested in \cite{Glavan:2019inb} that taking the limit~$D\!\rightarrow\!4$ after the rescaling of the Gauss-Bonnet coupling constant
\begin{equation}
\alpha \to \alpha/(D\!-\!4) \,
\label{coupling}
\end{equation}
can prevent vanishing of $H_{\mu\nu}$ in four dimensions.

The solution of (\ref{eq:EinsteinEq}) in the spherically  symmetric spacetime
can be written as
\begin{equation}
ds^{2}=-f(r)dt^{2}+\frac{1}{f(r)}dr^{2}+r^{2}(d\theta^{2}+\sin^{2}\theta d\phi^{2})\label{eq:metric}
\end{equation}
 with the metric function
\begin{equation}\label{sch-de}
f_\pm(r) = 1 + \frac{r^2}{\alpha }
	\left( 1\pm \sqrt{ \!1 \!+\! 2 \alpha \left( \frac{2 M}{r^3}+ \frac{\Lambda}{3}\right) } \!
	\right) \, ,
\end{equation}
where $M$ is a mass parameter. As the metric function $f_+(r)$ in the limit $\Lambda\rightarrow 0$ corresponds to asymptotically de Sitter case, here we will study $f_-(r)$, which is asymptotically flat when $\Lambda=0$.
Note that the black-hole metric (\ref{sch-de}) was considered earlier in \cite{Cognola,Cai:2009ua} in the context of the corrections to the entropy formula.

\begin{figure*}
\includegraphics[angle=0.0,width=0.5\linewidth]{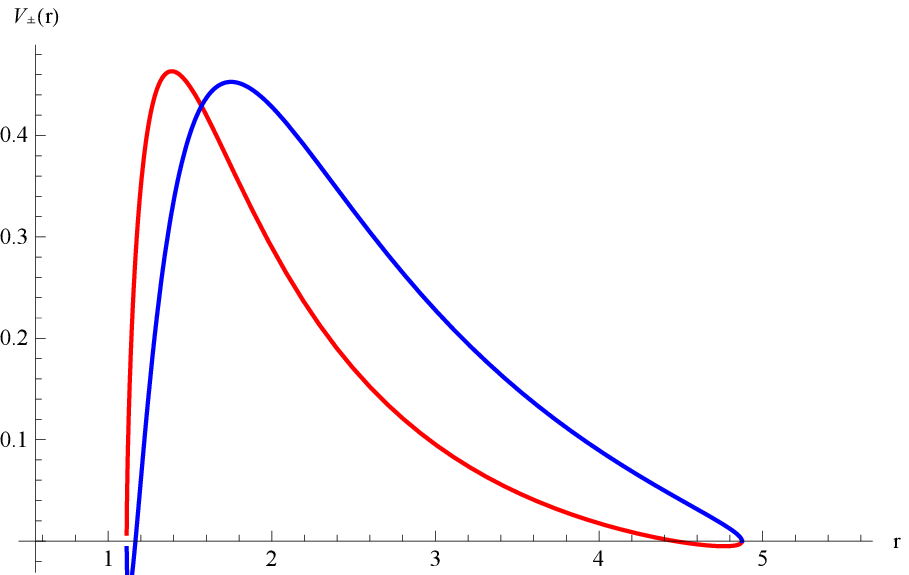}\includegraphics[angle=0.0,width=0.5\linewidth]{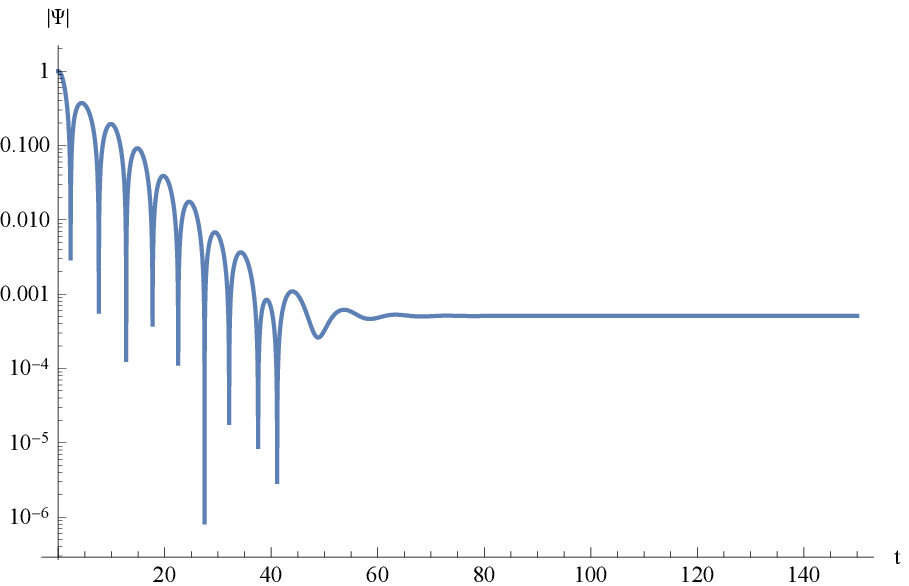}
\caption{Effective potentials $V_{+}$ (red) and $V_{-}$ (blue) (left panel) and the time domain profile (right panel) for the Dirac field with $\alpha=-0.15$, $\Lambda=0.1$, $k=2$, $M=\frac{1}{2}$.}
\label{fig1}
\end{figure*}

\begin{figure*}
\resizebox{\linewidth}{!}{\includegraphics*{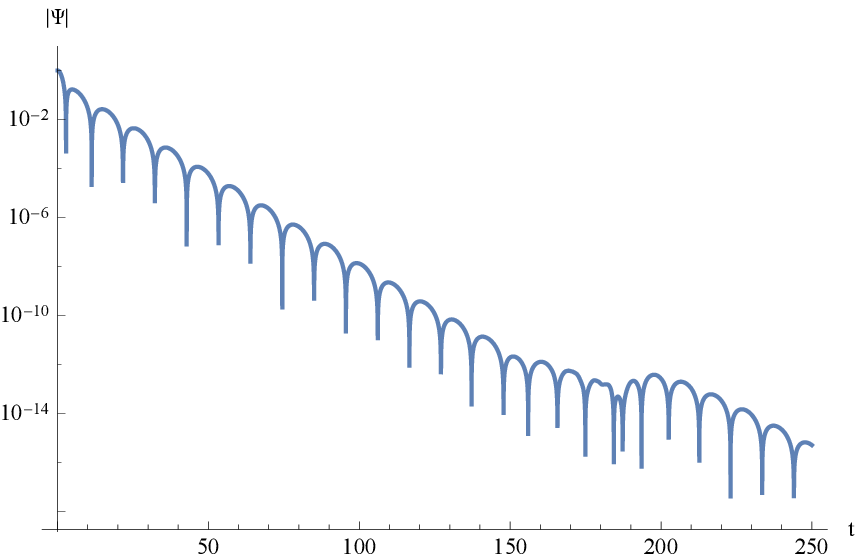}\includegraphics*{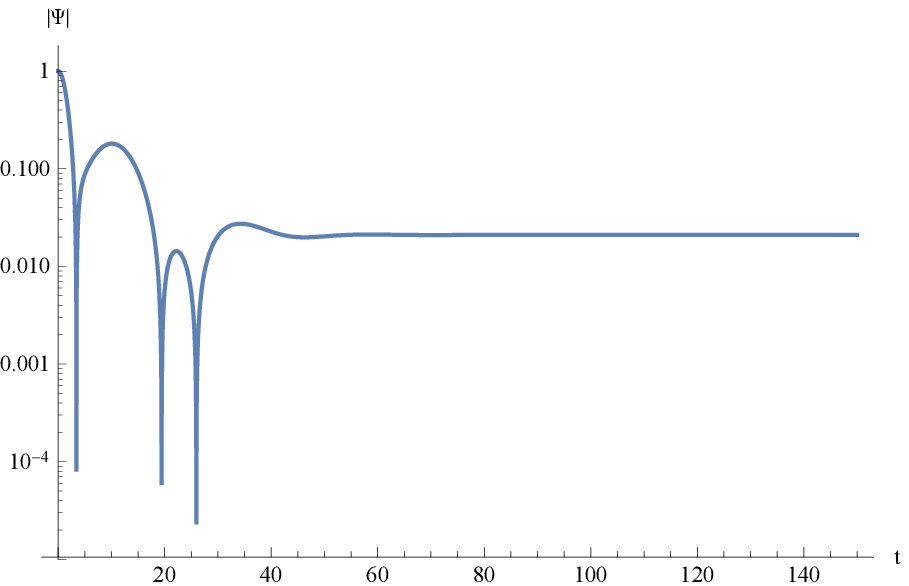}}
\caption{Examples of the time domain profile ($M=\frac{1}{2}$, $n=0$): with two concurrent modes for the electromagnetic field with $\Lambda=0.1$, $\alpha=-1.5$, $\ell=1$ (left panel) and for the Dirac field with $\Lambda=0.2$, $\alpha=-0.15$, $k=1$ (right panel).}\label{fig:TD1}
\end{figure*}

The metric (\ref{sch-de}), obtained via the regularisation scheme used in \cite{Glavan:2019inb}, is the solution of the consistent well-defined 4D Einstein-Gauss-Bonnet theory, formulated in \cite{Aoki:2020lig}.

\section{Master wave equations}

The general covariant equations for the test scalar $\Phi$, electromagnetic $A_\mu$ and Dirac $\Upsilon$ \cite{Brill:1957fx} fields have the form
\begin{equation}\label{KGg}
\frac{1}{\sqrt{-g}}\partial_\mu \left(\sqrt{-g}g^{\mu \nu}\partial_\nu\Phi\right)=0,
\end{equation}
\begin{equation}\label{EmagEq}
\frac{1}{\sqrt{-g}}\partial_\mu \left(\sqrt{-g}F^{\mu\nu}\right)=0\,,
\end{equation}
\begin{equation}\label{covdirac}
\gamma^{\alpha} \left( \frac{\partial}{\partial x^{\alpha}} - \Gamma_{\alpha} \right) \Upsilon=0,
\end{equation}
where $F_{\mu\nu}=\partial_\mu A_\nu-\partial_\nu A_\mu$, $\gamma^{\alpha}$ are noncommutative gamma matrices and $\Gamma_{\alpha}$ are spin connections in the tetrad formalism.
After separation of the variables Eqs. (\ref{KGg}) -- (\ref{covdirac}) take the following Schr\"{o}dinger-like form (see, for instance,  \cite{Konoplya:2011qq,Kokkotas:1999bd})
\begin{equation}\label{wave-equation}
\dfrac{d^2 \Psi}{dr_*^2}+(\omega^2-V(r))\Psi=0,
\end{equation}
where the "tortoise coordinate" $r_*$ is defined by the relation
\begin{equation}
dr_*=\frac{dr}{f(r)}.
\end{equation}

The effective potentials for the scalar ($s=0$) and electromagnetic ($s=1$) fields are
\begin{equation}\label{potentialScalar}
V(r)=f(r)\left(\frac{\ell\left(\ell+1\right)}{r^2}+\frac{1-s}{r}\cdot
\frac{d\,f(r)}{dr}\right),
\end{equation}
where $\ell=0, 1, 2, \ldots$ are the multipole numbers.
For the Dirac field we have two iso-spectral potentials
\begin{equation}
V_{\pm}(r) = \frac{k}{r}f(r) \left(\frac{k}{r}\mp\frac{\sqrt{f(r)}}{r}\pm \dfrac{d (\sqrt{f(r)})}{dr}\right),
\end{equation}
(where $k=1, 2, 3, \ldots$ are the multipole numbers) which can be transformed one into another by the Darboux transformation
\begin{equation}  \label{psi}
\Psi_{+}=q (W+\dfrac{d}{dr_*}) \Psi_{-}, \quad W=\sqrt{f(r)}, \quad q=const.
\end{equation}

If the effective potential $V$ is positive definite, the
differential operator
\begin{equation}
D = -\frac{\partial^{2}}{\partial r_{*}^2} + V
\end{equation}
is a positive self-adjoint operator in the Hilbert space of square integrable functions of $r_{*}$, and, therefore, all solutions of the perturbative equations of motion with
compact support initial conditions are bounded. This is the case of the scalar and electromagnetic perturbations, implying that the corresponding quasinormal spectrum has no growing modes indicating any kind of instability.

The proof of the Dirac field stability is non-trivial. For the asymptotically flat case ($\Lambda=0$) the potential $V_+(r)$ is positive definite, while the potential $V_-(r)$ has a negative gap near the event horizon. Still due to the iso-spectrality of the potentials the stability of the Dirac field for the other chirality is guaranteed \cite{Zinhailo:2019rwd}.

When the cosmological constant is turned on, the potential $V_+(r)$ also has a negative gap -- near the de Sitter horizon (see the left panel of the Fig. \ref{fig1}). In this situation stability of the Dirac field for the Schwarzschild-de Sitter black hole was proved analytically in \cite{LopezOrtega:2012hx} and using time domain integration in \cite{Konoplya:2020zso}. Here for the Einstein-Gauss-Bonnet-de Sitter spacetime we prove stability of the Dirac field with the help of the time domain integration, taking into account contributions of all the quasinormal modes. The obtained time domain profiles show the damped quasinormal oscillations approaching a static mode (an example is in the right panel of the Fig. \ref{fig1}).

The quasinormal modes are the solutions of the master wave equation (\ref{wave-equation}) satisfying the requirement of the purely outgoing waves at infinity and purely incoming waves at the event horizon (see,  for example, \cite{Konoplya:2011qq,Kokkotas:1999bd}).

We will be restricted by the values of the Gauss-Bonnet coupling constant $\alpha$ for which stability of the black hole is deliberately known, since there exists the phenomenon of the eikonal instability  \cite{Dotti:2005sq,Gleiser:2005ra,Konoplya:2017lhs,Konoplya:2017zwo,Takahashi:2010ye,
Yoshida:2015vua,Takahashi:2011qda,Gonzalez:2017gwa,Konoplya:2008ix,Cuyubamba:2016cug,Takahashi:2012np} of gravitational perturbations.

\section{Quasinormal modes}

We consider scalar, electromagnetic and Dirac perturbations of the 4D Einstein-Gauss-Bonnet-de Sitter black hole spacetime. For each of the test fields we fix $M=\frac{1}{2}$ and calculate the fundamental ($n=0$) quasinormal mode for the lower multipoles ($\ell=1$, $k=1$) as they dominate in the signal.

For calculation of the quasinormal modes we use the WKB method and the time domain integration. As both methods are well known and were recently surveyed in \cite{Konoplya:2019hlu,Konoplya:2011qq}, we briefly state their fundamentals.
\begin{enumerate}
\item In the frequency domain  we use the WKB method \cite{Schutz:1985zz}, extended to higher orders in \cite{Iyer:1986np,Konoplya:2003ii,Matyjasek:2017psv}.
We use the higher-order WKB formula \cite{Konoplya:2019hlu}:
$$ \omega^2=V_0+A_2(\K^2)+A_4(\K^2)+A_6(\K^2)+\ldots- $$
\begin{equation}\label{HighWKB}
\imo \K\sqrt{-2V_2}\left(1+A_3(\K^2)+A_5(\K^2)+A_7(\K^2)\ldots\right),
\end{equation}
where $\K=\textrm{sign}\re\omega\left(n+\frac{1}{2}\right)$, $n=0,1,2,3\ldots$. The corrections $A_k(\K^2)$ of order $k$ to the eikonal formula are polynomials of $\K^2$ with rational coefficients and depend on the values of higher derivatives of the potential $V(r)$ in its maximum. We also use Padé approximants \cite{Matyjasek:2017psv,Hatsuda:2019eoj} to increase accuracy of the higher-order WKB formula (\ref{HighWKB}).

\item In the time domain we use integration of the wave equation (without the stationary ansatz) at a fixed point in the space \cite{Gundlach:1993tp}.
We integrate the wave-like equation rewritten in terms of the light-cone variables $u=t-r_*$ and $v=t+r_*$. The appropriate discretization scheme was suggested in \cite{Gundlach:1993tp}:
$$
\Psi\left(N\right)=\Psi\left(W\right)+\Psi\left(E\right)-\Psi\left(S\right)-
$$
\begin{equation}\label{Discretization}
-\Delta^2\frac{V\left(W\right)\Psi\left(W\right)+V\left(E\right)\Psi\left(E\right)}{8}+{\cal O}\left(\Delta^4\right)\,,
\end{equation}
where we used the following notation for the points:
$N=\left(u+\Delta,v+\Delta\right)$, $W=\left(u+\Delta,v\right)$, $E=\left(u,v+\Delta\right)$ and $S=\left(u,v\right)$. The initial data are given on the null surfaces $u=u_0$ and $v=v_0$.
\end{enumerate}

\begin{table}
\begin{tabular}{|c|c|c|}
  \hline
  $\alpha$ & $QNM$ (WKB) & $QNM$ (Time-domain) \\
  \hline
   \multicolumn{3}{|c|}{$\Lambda=0$}\\
  \hline
  $-1.50$ &$0.484276 - 0.254452 i$ & $0.473516 - 0.247996 i$ \\
  $-0.90$ &$0.515453 - 0.236491 i$ & $0.514174 - 0.234622 i$ \\
  $-0.15$ &$0.571781 - 0.204173 i$ & $0.574111 - 0.202620 i$ \\
  $0.15$ & $0.601189 - 0.184841 i$ & $0.601748 - 0.183514 i$\\
  \hline
  \multicolumn{3}{|c|}{$\Lambda=0.1$}\\
  \hline
  $-1.50$ &$0.344319 - 0.209063 i$ & $0.354374 - 0.201506 i$ \\
  $-0.90$ &$0.400102 - 0.201211 i$ & $0.405111 - 0.199318 i$ \\
  $-0.15$ &$0.483659 - 0.183967 i$ & $0.484016 - 0.183977 i$ \\
  $0.15$ & $0.524844 - 0.170221 i$ & $0.524841 - 0.170066 i$\\
  \hline
  \multicolumn{3}{|c|}{$\Lambda=0.2$}\\
  \hline
  $-1.50$ &$0.178574 - 0.106481 i$ & $0.179645 - 0.106425 i$ \\
  $-0.90$ &$0.263647 - 0.138567 i$ & $0.264520 - 0.138511 i$ \\
  $-0.15$ &$0.383315 - 0.153947 i$ & $0.383321 - 0.154055 i$ \\
  $0.15$ & $0.441451 - 0.149673 i$ & $0.441473 - 0.149627 i$\\
  \hline
\end{tabular}
\caption{Scalar fundamental quasinormal mode, calculated by WKB and time-domain methods, for various values of the coupling constant $\alpha$ in the stability sector; $\ell=1$, $n=0$, $M=\frac{1}{2}$.}
\end{table}

\begin{table}
\begin{tabular}{|c|c|c|}
  \hline
  $\alpha$ & $QNM$ (WKB) & $QNM$ (Time-domain) \\
  \hline
   \multicolumn{3}{|c|}{$\Lambda=0$}\\
  \hline
  $-1.50$ &$0.378701 - 0.222768 i$ & $0.368476 - 0.216081 i$ \\
  $-0.90$ &$0.413577 - 0.214727 i$ & $0.410388 - 0.212675 i$ \\
  $-0.15$ &$0.479068 - 0.192548 i$ & $0.479512 - 0.192325 i$ \\
  $0.15$ & $0.515695 - 0.175396 i$ & $0.515753 - 0.174584 i$\\
  \hline
  \multicolumn{3}{|c|}{$\Lambda=0.1$}\\
  \hline
  $-1.50$ &$0.294169 - 0.178994 i$ & $0.297708 - 0.171592 i$ \\
  $-0.90$ &$ 0.340064 - 0.180570 i$ & $0.342966 - 0.175107 i$ \\
  $-0.15$ &$0.419706 - 0.168264 i$ & $0.419886 - 0.168088 i$ \\
  $0.15$ & $0.462478 - 0.157339 i$ & $0.462448 - 0.157328 i$\\
  \hline
  \multicolumn{3}{|c|}{$\Lambda=0.2$}\\
  \hline
  $-1.50$ &$0.170134 - 0.09743 i$ & $0.171873 - 0.097341 i$ \\
  $-0.90$ &$0.242860 - 0.123987 i$ & $0.244510 - 0.123562 i$ \\
  $-0.15$ &$0.347241 - 0.139301 i$ & $0.347504 - 0.139189 i$ \\
  $0.15$ & $0.401234 - 0.136861 i$ & $0.401206 - 0.136798 i$\\
  \hline
\end{tabular}
\caption{Electromagnetic fundamental quasinormal mode, calculated by WKB and time-domain methods, for various values of the coupling constant $\alpha$ in the stability sector; $\ell=1$, $n=0$, $M=\frac{1}{2}$.}
\end{table}

\begin{table}
\begin{tabular}{|c|c|c|}
  \hline
  $\alpha$ & $QNM$ (WKB) & $QNM$ (Time-domain) \\
  \hline
   \multicolumn{3}{|c|}{$\Lambda=0$}\\
  \hline
  $-1.50$ &$0.271919 - 0.304772 i$ & $0.272825 - 0.302579 i$ \\
  $-0.90$ &$0.315778 - 0.252119 i$ & $0.312684 - 0.262944 i$ \\
  $-0.15$ &$0.355619 - 0.204282 i$ & $0.358747 - 0.205731 i$ \\
  $0.15$ & $0.376604 - 0.181891 i$ & $0.381600 - 0.180157 i$\\
  \hline
  \multicolumn{3}{|c|}{$\Lambda=0.1$}\\
  \hline
  $-1.50$ &$0.237188 - 0.204896 i$ & $0.244334 - 0.202384 i$ \\
  $-0.90$ &$0.267136 - 0.194608 i$ & $0.266835 - 0.197179 i$ \\
  $-0.15$ &$0.313254 - 0.175431  i$ & $0.337239 - 0.173562 i$ \\
  $0.15$ & $0.339161 - 0.161514 i$ & $0.354837 - 0.153228 i$\\
  \hline
  \multicolumn{3}{|c|}{$\Lambda=0.2$}\\
  \hline
  $-1.50$ &$0.139785 - 0.102606 i$ & $0.137811 - 0.108875 i$ \\
  $-0.90$ &$0.192369 - 0.129557 i$ & $0.2002824 - 0.146257 i$ \\
  $-0.15$ &$0.260935 - 0.142754 i$ & $0.279983 - 0.146596 i$ \\
  $0.15$ & $0.295795 - 0.139156 i$ & $0.316639 - 0.135680 i$\\
  \hline
\end{tabular}
\caption{Dirac fundamental quasinormal mode, calculated by WKB and time-domain methods, for various values of the coupling constant $\alpha$ in the stability sector; $k=1$, $n=0$, $M=\frac{1}{2}$.}
\end{table}

\begin{figure*}
\resizebox{\linewidth}{!}{\includegraphics*{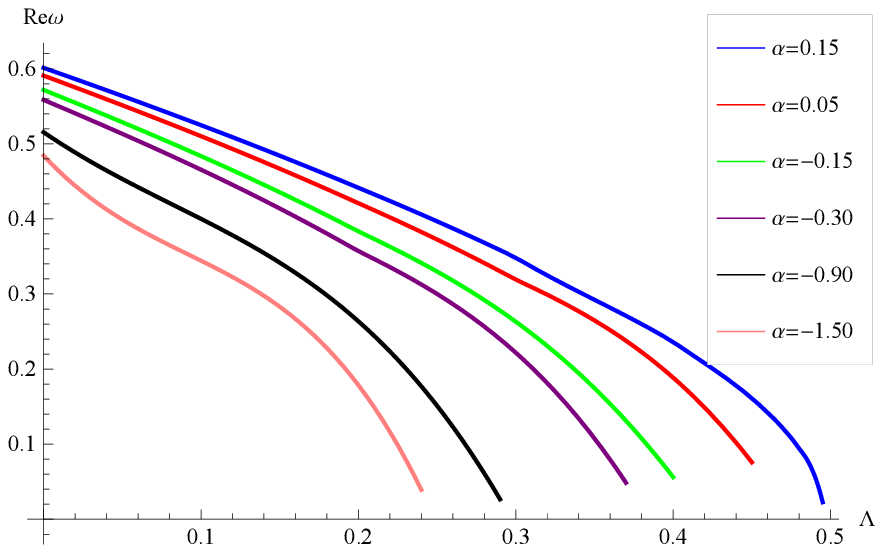}\includegraphics*{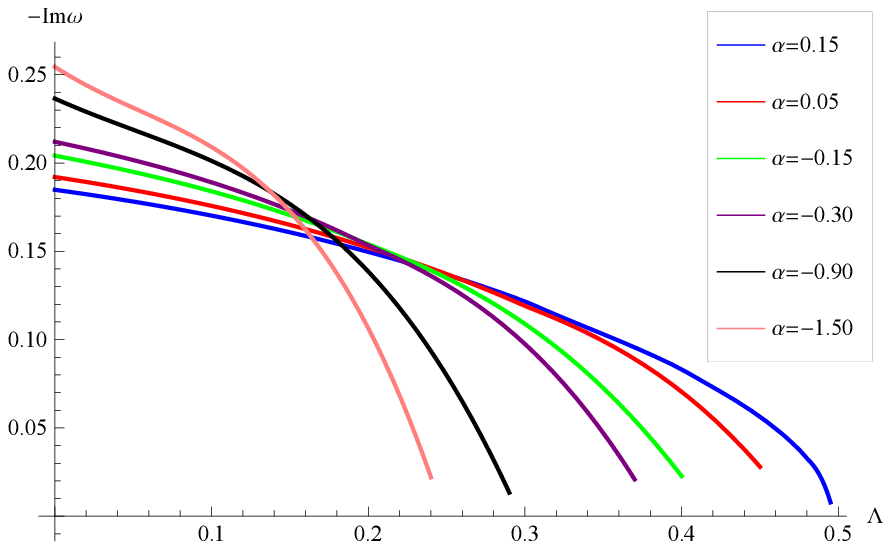}}
\caption{Real (left panel) and imaginary (right panel) part of the scalar fundamental quasinormal mode, calculated by the WKB method, depending on $\Lambda$, for various values of the coupling constant $\alpha$ in the stability sector; $\ell=1$, $s=0$, $n=0$, $M=\frac{1}{2}$.}\label{fig:ScalarL1}
\end{figure*}

\begin{figure*}
\resizebox{\linewidth}{!}{\includegraphics*{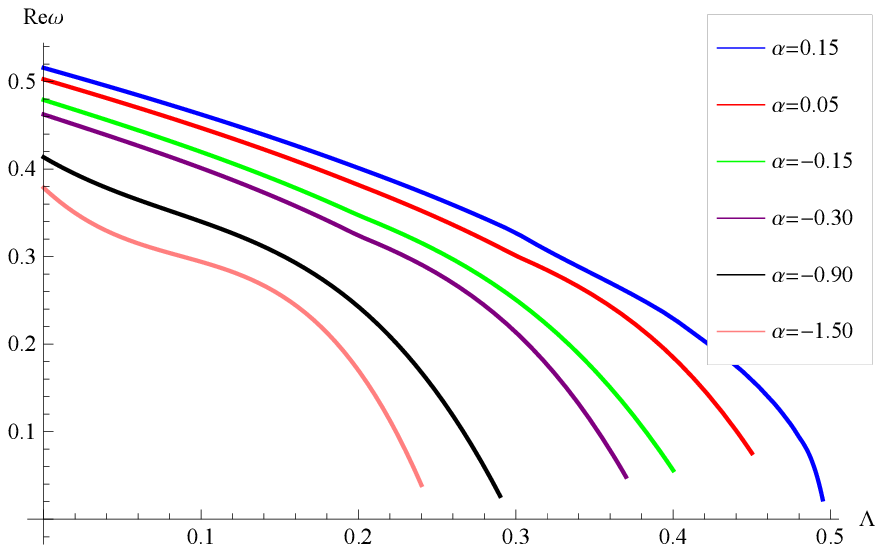}\includegraphics*{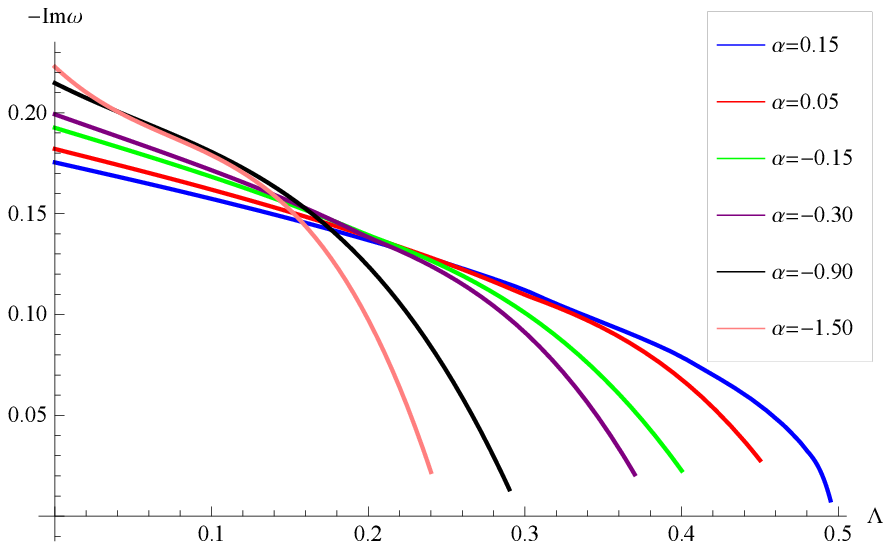}}
\caption{Real (left panel) and imaginary (right panel) part of the electromagnetic fundamental quasinormal mode, calculated by the WKB method, depending on $\Lambda$, for various values of the coupling constant $\alpha$ in the stability sector; $\ell=1$, $s=1$, $n=0$, $M=\frac{1}{2}$.}\label{fig:EmagL1}
\end{figure*}

\begin{figure*}
\resizebox{\linewidth}{!}{\includegraphics*{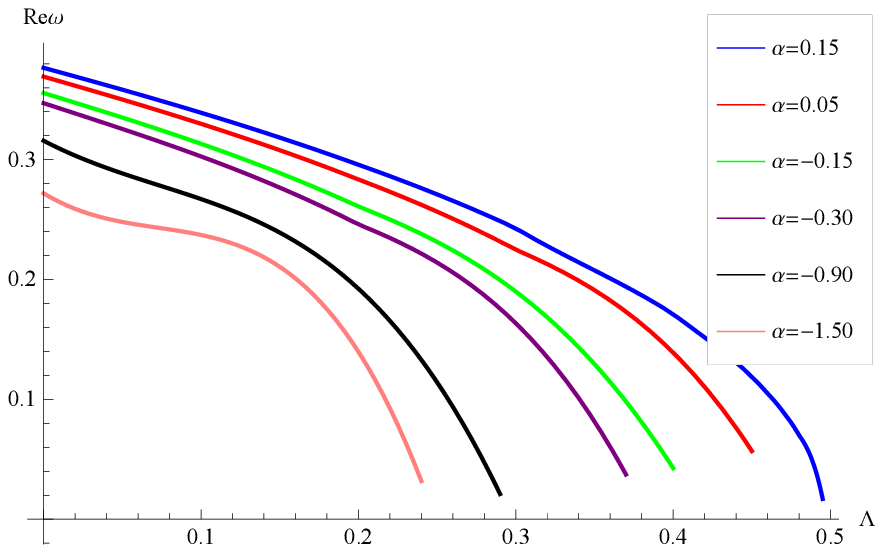}\includegraphics*{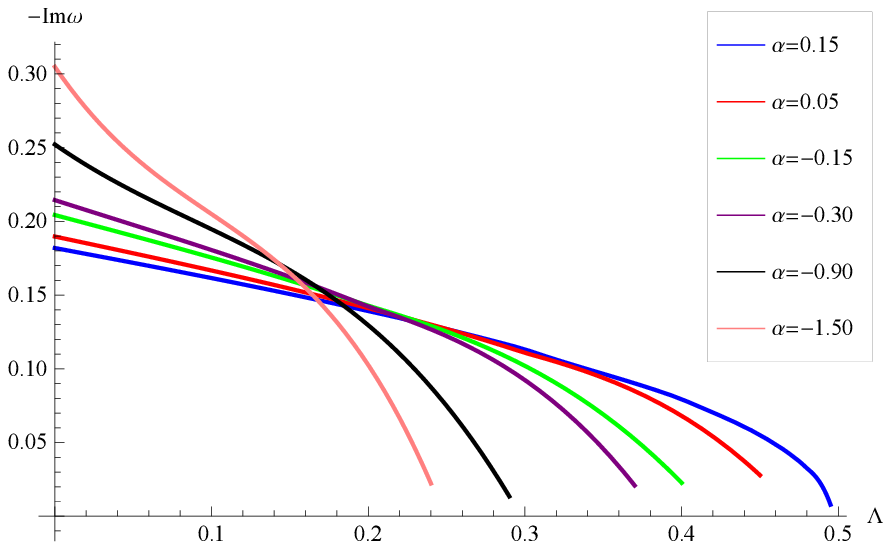}}
\caption{Real (left panel) and imaginary (right panel) part of the Dirac fundamental quasinormal mode, calculated by the WKB method, depending on $\Lambda$, for various values of the coupling constant $\alpha$ in the stability sector; $k=1$, $n=0$, $M=\frac{1}{2}$.}\label{fig:Dirack1}
\end{figure*}

We used for the most part the sixth order WKB method with Pad\'{e} approximants \cite{Matyjasek:2017psv} (taking $\tilde{n} =2$, $\tilde{m} =4$ for the scalar and electromagnetic fields and $\tilde{n} =1$, $\tilde{m} =5$ for the Dirac field, where $\tilde{n},\;\tilde{m}$ are defined in  \cite{Konoplya:2019hlu} and denote the orders of the polynomials in the nominator and denominator correspondingly) and the time-domain integration to check the obtained results. It can be seen from the Tables I -- III that the values of the quasinormal modes calculated by both methods in the stability sector for the coupling constant $\alpha$ are in a good correspondence. At $\alpha \approx -1.5$ and less the agreement between the WKB method and the time-domain integration is somewhat worse, which can be explained by appearance of the two concurrent modes in the time-domain profile (see an example in the left panel of the Fig. \ref{fig:TD1}). The discrepancy between the results obtained by the WKB method and the time domain integration for the Dirac field (see Table III) is caused by the difficulty of extracting the quasinormal mode out of the time domain profile, which has only a few oscillations for the lower multipole (an example is in the right panel of the Fig. \ref{fig:TD1}). Note that for the de Sitter case the late-time tail of the time domain profile for the Dirac field has a specific shape \cite{Konoplya:2020zso}, approaching a horizontal line, which points out the presence of the static zero mode.

The real and imaginary parts of the fundamental ($n=0$) quasinormal mode, calculated by the WKB method for various values of the coupling constant $\alpha$ in the stability sector \cite{Konoplya:2020bxa}, depending on the cosmological constant $\Lambda$, are presented in Figs. \ref{fig:ScalarL1} -- \ref{fig:Dirack1}. The presence of the cosmological constant $\Lambda$ in the metric function (\ref{sch-de}) implies that in addition to the event horizon there appears the second, de Sitter horizon. For every fixed value of $\alpha$ in response to the increasing of $\Lambda$ both horizons are closing in until we reach some extremal value of the cosmological constant. Therefore the plots in Figs. \ref{fig:ScalarL1} -- \ref{fig:Dirack1} are of different length in $\Lambda$, at that the less $\alpha$ is, the earlier the extremal value of $\Lambda$ is reached.

It can be seen that increasing of the cosmological constant suppresses both the real oscillation frequency and the damping rate at any value of the Gauss-Bonnet coupling constant. As to the changes of the coupling constant $\alpha$, the corresponding behaviour of the quasinormal modes is qualitatively similar to that in the Schwarzschild case \cite{Konoplya:2020bxa,Churilova:2020aca}: the damping rate is more sensitive to increasing of the coupling constant, while the real oscillation frequency is monotonically increased.

\section{Conclusions}

The black hole solutions, obtained as a result of the regularization scheme used in \cite{Glavan:2019inb}, turned out to be the solutions of the consistent well-defined $4$-dimensional Einstein-Gauss-Bonnet theory of gravity formulated in \cite{Aoki:2020lig}.
Recently the quasinormal modes of the bosonic \cite{Konoplya:2020bxa} and fermionic \cite{Churilova:2020aca} fields for the spherically symmetric asymptotically flat black hole in this theory \cite{Aoki:2020lig} were studied. We extended these results to the de Sitter case of this novel theory of gravity by calculation of the quasinormal modes of the test scalar, electromagnetic and Dirac fields for the spherically symmetric black hole.

We have found that increasing of the cosmological constant suppresses both the real oscillation frequency and the damping rate at any value of the Gauss-Bonnet coupling constant. As to the changes of the coupling constant $\alpha$, they affect the damping rate to a greater extent than the real oscillation frequency, which is monotonically increased.

While the stability of the scalar and electromagnetic fields follows from the positive definiteness of the effective potential, there is no such positive definiteness for the Dirac field \cite{LopezOrtega:2012hx,Konoplya:2020zso}. We proved stability of the Dirac field in $4D$ Einstein-Gauss-Bonnet-de Sitter theory, using the time domain integration, which takes into account contributions of all the quasinormal modes.

Our work can be further extended by studying of the gravitational perturbations in this novel Einstein-Gauss-Bonnet-de Sitter theory.

\textit{Note added.} When we were preparing our paper for publication, a work \cite{Liu:2020evp} appeared, which considers quasinormal modes of the scalar charged field, what has a rather small overlap with our paper in the part considering scalar field modes.

\acknowledgments{
The author acknowledges Roman Konoplya for useful discussions and Alexander Zhidenko for kind help.}

\end{document}